# Fuzzy cellular model of signal controlled traffic stream

*Bartłomiej Płaczek*

Faculty of Transport, Silesian University of Technology,
Krasińskiego 8, 40-019 Katowice, Poland
bartlomiej.placzek@polsl.pl

Abstract: Microscopic traffic models have recently gained considerable importance as a mean of optimising traffic control strategies. Computationally efficient and sufficiently accurate microscopic traffic models have been developed based on the cellular automata theory. However, the real-time application of the available cellular automata models in traffic control systems is a difficult task due to their discrete and stochastic nature. This paper introduces a novel method of traffic streams modelling, which combines cellular automata and fuzzy calculus. The introduced fuzzy cellular traffic model eliminates main drawbacks of the cellular automata approach i.e. necessity of multiple Monte Carlo simulations and calibration issues. Experimental results show that the evolution of a simulated traffic stream in the proposed fuzzy cellular model is consistent with that observed for stochastic cellular automata. The comparison of both methods confirms that the computational cost of traffic simulation is considerably lower for the proposed model. The model is suitable for real-time applications in traffic control systems.

## 1. Introduction

The development of adequate traffic models for applications in the road traffic control is a challenging issue. Such models should present a well-balanced trade-off between accuracy and computational complexity to enable on-line processing of measurement data and traffic state estimation. The state-of-the-art traffic control methods use macroscopic and mesoscopic models that describe queues or groups of vehicles [1, 2, 3]. However, the individual features related to vehicles are important from the traffic control point of view, as they have a significant influence on the traffic performance.

Modern sensing platforms (e.g. vision-based monitoring systems [4, 5] and vehicular sensor networks [6]) offer traffic data concerning the parameters of particular vehicles (position, velocity, acceleration, direction, etc.). The vehicles can be used as sources of information to determine detailed traffic stream characteristics. Emerging technologies in the road traffic monitoring enable wireless communication between sensing devices installed in vehicles (mobile sensors) and the road environment for dynamic transfers of measurement data [7]. These data cannot be fully utilised for traffic control purposes when using macroscopic or mesoscopic models.

Micro-simulation as a mean of controlling traffic systems has recently gained the considerable importance [8]. Computationally efficient and sufficiently accurate microscopic traffic models have been developed based on the cellular automata theory [9]. However, application of the available cellular automata models in traffic control systems is a difficult task due to their stochastic character. Stochastic parameters in the cellular automata are necessary to represent model uncertainties and to enable a model calibration. Unfortunately, the traffic simulation with stochastic cellular automata requires the time consuming Monte Carlo method to be used. Long computational time of the Monte Carlo simulation is a critical disadvantage for traffic control applications that require the results of simulation to be obtained in a strictly limited time-frames.

In this paper a new microscopic traffic model is introduced, which does not involve the Monte Carlo technique and enables a realistic simulation of signal controlled traffic streams. The model was formulated as a hybrid system combining a fuzzy calculus with the cellular automata approach. The original feature distinguishing this model from the other cellular models is that vehicle position, its velocity and other parameters are modelled by fuzzy numbers. Moreover, the rule of model transition from one time step to the next is also based on fuzzy definitions of basic arithmetic operations.

The application of fuzzy calculus helps to deal with imprecise traffic data and to describe uncertainty of the simulation results. In fact, it is impossible to predict unambiguously the effect of an application of a specific traffic control strategy. Moreover, the current traffic state also cannot be usually identified precisely on the basis of the available measurement data. Therefore, the fuzzy numbers are used in a traffic model to describe the uncertainty and precision of the simulation inputs and outputs. The model allows a single simulation to take into account many potential scenarios (traffic state evolutions) [10]. These facts along with low computational complexity make the model suitable for real-time applications in traffic control systems.

The rest of the paper is organised as follows: Related works are reviewed and analysed in Section 2. Section 3 addresses limitations regarding the use of cellular automata for modelling the traffic at signalised intersections. Section 4 introduces the fuzzy cellular model of signalised traffic stream and describes the traffic simulation algorithm in details. A comparison of simulation results for the fuzzy cellular model and the Nagel-Schreckenberg stochastic cellular automata is presented in Section 5. Finally, conclusions are given in Section 6.

## 2. Related works

Cellular automata have become a frequently used tool for microscopic modelling of road traffic processes. Their main advantages are high computational efficiency and fast performance when used in computer simulations. The cellular automata traffic models are dynamical systems with discrete time, space and state variables. Despite limited accuracy on a microscopic scale, they allow real traffic phenomena to be simulated with sufficient precision. A comprehensive and detailed review of the cellular automata traffic models can be found in [11, 12].

Among many applications in the field of road traffic modelling, the cellular automata were also used for simulation and optimisation of signal traffic control. In [13] a traffic simulation tool for urban road networks was proposed which is based on Nagel-Schreckenberg (NaSch) stochastic cellular automata [14]. An intersection model was considered in this work including traffic regulations (priority rules, signs, and signalisation). It was also suggested that for appropriate setting of a deceleration probability parameter the model yields realistic time headways between vehicles crossing stop line at a signalised intersection.

A modified NaSch model for traffic flow controlled by a traffic signal was proposed in [15]. According to the introduced modification, the deceleration probability for each vehicle is determined as a function of free space in front of the vehicle. The model was applied in order to simulate a signal controlled traffic flow on a single-lane road. Several models of this type that are based on the NaSch cellular automata can be found in the literature (e.g.: model with turning-deceleration rule [16], model with anticipation of change in traffic lights [17]).

Schadschneider et al. [18] have presented a cellular automata model of vehicular traffic in signalised urban networks by combining ideas borrowed from Biham-Middleton-Levine model of city traffic [19] and the NaSch model of single lane traffic stream. The similar model was adopted to calculate optimal parameters of traffic signal coordination plan that maximise a flow in a road network [20].

In [21] a model of city traffic was introduced, which is based on deterministic elementary cellular automata. Each cell in an elementary cellular automaton has only two possible states (0 and 1). Moreover, the state of a cell depends only on the present states of its nearest neighbours. The simplicity of this model allows for the simulation of large road networks with many intersections. The emphasis in this approach was put on the simplicity and scalability of the model rather than on realism of the traffic simulation.

A cellular automata traffic model was also utilised as an evaluation tool in a genetic algorithm for the traffic signals optimisation [9]. The fitness function in this algorithm is evaluated on the basis of traffic simulation results. The optimisation was performed for a road network with 20 signalised intersections. The results were compared for a stochastic and a deterministic version of the cellular automata model. It was observed that the obtained population fitness ranking is similar for both versions. However, the deterministic cellular automata have enabled a remarkable speed-up of the genetic algorithm execution.

The relationships between parameters of the cellular automata models and saturation flow rates at simulated intersection were analysed in [22]. The traffic models were investigated in this work to identify the possibility of reproducing any desired value of the saturation flow. This analysis was performed for both the deterministic and stochastic cellular automata. It was concluded that the stochastic version allows any value of the saturation flow to be obtained by adjusting a deceleration probability parameter.

Various artificial intelligence techniques have been used in the field of traffic modelling [23, 24]. In this paper a cellular automata model of road traffic is combined with fuzzy arithmetic. Hybrid artificial intelligence systems that combine the cellular automata and fuzzy sets are typically referred to as fuzzy cellular automata (FCA) [25]. FCA-based models have found many applications in the field of complex systems simulation e.g. [26, 27]. A road traffic model of this kind has been proposed in [28]. In such models, the local update rule of classical cellular automata is usually replaced by a fuzzy logic system consisting of fuzzy rules, fuzzification, inference, and defuzzification mechanisms. A different approach is used in this paper: current states of the cells are determined by fuzzy sets and a calculus with fuzzy numbers is involved in the update operation. The innovative features of the proposed methodology are the elimination of information loss caused by defuzzification and the incorporation of uncertainty in simulation results.

## 3. Limitations of cellular automata models

Cellular automata models of road traffic describe velocities and positions of vehicles in discrete time steps. Position $x_{i,t}$ indicates a cell, which is occupied by vehicle $i$ at time step $t$. Velocity $v_{i,t}$ is expressed in cells per time step and determines how many cells the vehicle $i$ advance at time step $t$. The discrete positions and velocities are updated at each time step according to the rule of the cellular automata. In order to compute velocity, the rule takes into account previous velocity values, a maximal velocity $v_{max}$ and number of free cells in front of the vehicle $i$ at time step $t$ (so-called gap) $g_{i,t}$.

Realistic simulation of the signalised intersection requires a traffic model, which can be appropriately calibrated to reflect real values of the measured saturation flow i.e. the maximum hourly vehicle flow rate, at which the traffic is as dense as could reasonably be expected, passing an intersection under prevailing roadway, traffic, and control conditions. For cellular automata traffic models the calibration is not a trivial task due to their discrete formulation and limited set of parameters.

## 3.1. Deterministic cellular automata

In case of deterministic cellular automata models the saturation flow rate can be evaluated by the analysis of queue discharge behaviour. To this end a traffic stream have to be modelled, which consists of vehicles leaving a queue after end of red signal. In such traffic stream a uniform gap $g$ exists between vehicles that reached the maximal velocity $v_{max}$. Every vehicle occupies one cell, thus the gap $g$ corresponds to the traffic density of $1/(g+1)$ vehicles per cell. On this basis the saturation flow $s$ can be calculated in vehicles per time step:

$$s = \frac{v_{max}}{g+1}. \tag{1}$$

According to the above equation, the saturation flow in a deterministic cellular automata model can be adjusted in two ways: by changing the maximal velocity or changing the gap that occurs between vehicles moving with maximal velocity. The latter requires a modification of the cellular automata rule. In practice, it is not possible to obtain an arbitrary saturation flow rate because $v_{max}$ and $g$ takes only integer values and a set of applicable rules that reproduce the real-life traffic behaviour is very limited.

In order to illustrate the issue of deterministic model calibration, saturation flow rate will be analysed taking into account three different cellular automata rules. Fig. 1 compares the queue discharge behaviour for the three considered rules. In this example, the maximal velocity is two cells per time step. Numbers in Fig. 1 denote velocities of vehicles and indicate their positions (occupied cells), the symbol "X" represents a red signal for vehicles, which is active only at the first time step of the simulation.

The first rule (R1) is similar to the rule that was proposed by Takayasu and Takayasu [29]. According to this model a stopped vehicle starts to move only if the gap in front of it $g_{i,t}$ is wider than one cell:

$$v_{i,t} = \begin{cases} 0, & v_{i,t-1} = 0 \land g_{i,t} = 1, \\ \min(v_{i,t-1}+1, g_{i,t}, v_{max}), & \text{else,} \end{cases} \tag{2}$$

$$x_{i,t+1} = x_{i,t} + v_{i,t}.$$

When applying the above formulas, the resulting gap $g$ between vehicles in saturated traffic stream is of 2 $v_{max}$ cells (Fig. 1 a).

A smaller gap $g$ is obtained for rule R2, which skips the update of a vehicle position if the vehicle is stopped and its current gap $g_{i,t}$ is of one cell:

$$v_{i,t} = \min(v_{i,t-1}+1, g_{i,t}, v_{max}), \tag{3}$$

$$x_{i,t} = \begin{cases} x_{i,t}, & v_{i,t-1} = 0 \land g_{i,t} = 1, \\ x_{i,t} + v_{i,t}, & \text{else.} \end{cases}$$

Using rule R2 a saturated traffic stream is formed with two gap widths: $g = v_{max} + 1$ for vehicles with even indexes and $g = 2\ v_{max} - 1$ for vehicles with odd indexes. Note that the average gap is 1.5 $v_{max}$ and both above values are equal if $v_{max} = 2$ (Fig. 1 b).

The third analysed rule (R3) corresponds to the deterministic case of the Nagel-Schreckenberg model [14] with slowdown probability parameter $p = 0$:

$$v_{i,t} = \min(v_{i,t-1}+1, g_{i,t}, v_{max}), \tag{4}$$

$$x_{i,t+1} = x_{i,t} + v_{i,t}.$$

For this rule the gap $g$ has width of $v_{max}$ cells (Fig. 1 c).

Gaps $g$ and the resulting saturation flow rates for all above rules are summarised in Tab. 1. The saturation flow rates $s$ in vehicles per hour of green time were calculated assuming that $v_{max} = 2$ and one time step corresponds to one second.

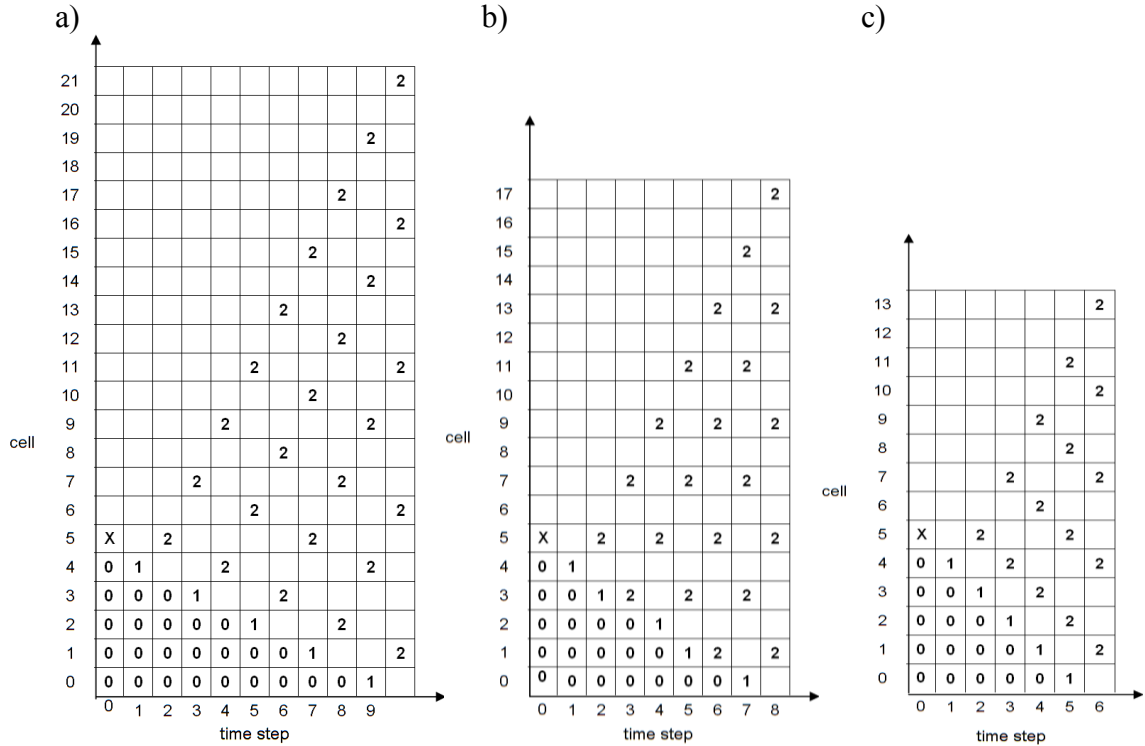

Fig. 1. Queue discharge behaviour for deterministic cellular automata rules R1-R3

Tab. 1. Saturation flow rates for deterministic cellular automata rules R1-R3

| Rule | $g$ [cells] | $s$ [vehs per time step] | $s$ [vehs per hour] |
|------|-------------|--------------------------|---------------------|
| R1   | $2v_{max}$  | $\dfrac{v_{max}}{2v_{max}+1}$ | 1440 |
| R2   | $1.5v_{max}$ | $\dfrac{v_{max}}{1.5v_{max}+1}$ | 1800 |
| R3   | $v_{max}$   | $\dfrac{v_{max}}{v_{max}+1}$ | 2400 |

The presented examples show that it is possible to obtain only a limited set of saturation flow rates by manipulating parameters and rules of the deterministic cellular automata models. Moreover, minimal model modifications result in significant changes of the saturation flow. Thus, the deterministic cellular automata models are not sufficient for the realistic traffic simulation at signalised intersections.

### 3.2. Stochastic cellular automata

Stochastic cellular automata models include some additional probability parameters that control the random aspects of traffic simulation. As it was discussed in [22], the probability parameters have a direct influence on saturation flow rates of the simulated traffic stream. However, the modification of the probability parameter for cellular automata model results not only in change of the average (expected) value of the saturation flow rates, but also in change of their spread. This effect is illustrated in Figs. 2 and 3 for the NaSch model.

The NaSch model was used to simulate traffic at a signalised intersection. During the experiment the deceleration probability parameter $p$ was changed between 0 and 0.8 with increments of 0.01. The simulation of one hour period was repeated five hundred times for every value of the probability $p$. Saturation flow rate was calculated in each simulation run. On this basis a distribution of the saturation flow rates was determined for every value of parameter $p$ in the analysed range. The plot in Fig. 2 shows medians, 5-th, and 95-th percentiles of the obtained saturation flow distributions. An example of the distribution histogram for $p = 0.2$ is presented in Fig. 6. The spread of the saturation flow rates was evaluated as a difference between 95-th and 5-th percentile (Fig. 3).

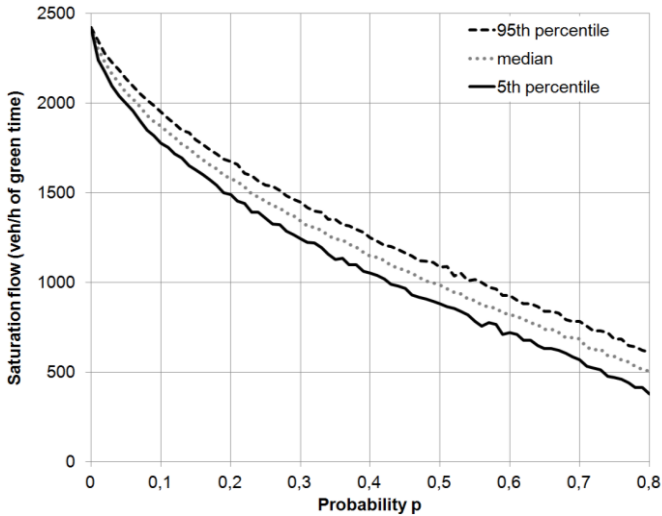

Fig. 2. Saturation flow rate vs. probability parameter $p$ for NaSch model

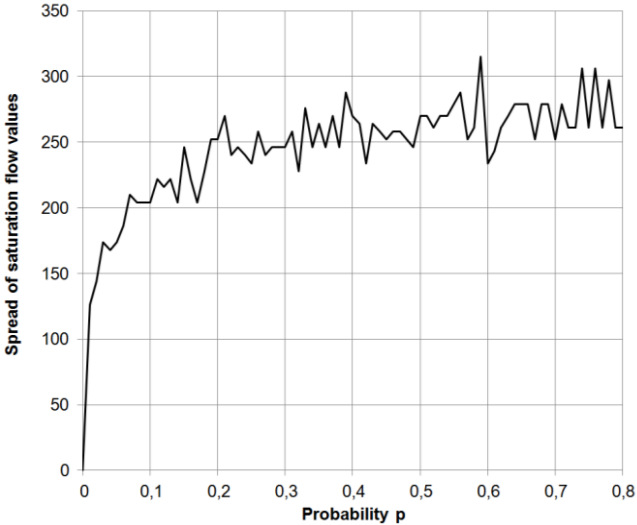

Fig. 3. Spread of saturation flow values vs. probability parameter $p$ for NaSch model

From the results in Figs. 2 and 3 it may be concluded that an increase of probability $p$ causes both lower saturation flow rates and higher spread of its values. Thus, for the NaSch model it is impossible to independently change the average value and the range of the saturation flow rates. Another issue is related to the dependency between free-flow velocity $v_f$ and probability parameter: $v_f = v_{max} - p$, which is a well-known characteristic of the NaSch cellular automata [11]. Due to this dependency, any modification of the probability parameter results also in the change of the free-flow traffic velocity. The aforementioned mutual dependencies between parameters in the stochastic cellular automata seriously impede the use of the probability parameter for traffic model calibration.

Application of stochastic cellular automata for the evaluation of traffic performance at a signalised intersection requires the Monte Carlo method to be used for estimation of performance measures [9]. A number of traffic simulation runs is necessary to establish a meaningful estimate. Therefore the applicability of the stochastic cellular automata is limited due to the long computational time of the Monte Carlo simulation. This disadvantage is critical in traffic control applications, where the results have to be obtained faster than the real duration time of the simulated process.

## 4. Fuzzy cellular model

A fuzzy cellular model of road traffic was developed to overcome the limitations of cellular automata models that were discussed in previous section. The introduced model combines the main advantages of cellular automata models with a possibility of realistic traffic simulation at signalised intersections. The proposed method allows the traffic model to be calibrated in order to reflect real values and uncertainties of measured saturation flows.

A traffic lane in the fuzzy cellular model is divided into cells that correspond to the road segments of equal length. The traffic state is described in discrete time steps. These two basic assumptions are consistent with those of the Nagel-Schreckenberg cellular automata model. Thus, the methods presented in [11] are also applicable here for the determination of cell length and vehicles properties. A novel feature in this approach is that vehicle parameters are modelled using ordered fuzzy numbers [30]. Moreover, the model transition from one time step ($t$) to the next ($t + 1$) is based on fuzzy definitions of basic arithmetic operations.

The road traffic stream is represented in the fuzzy cellular model as a set of vehicles. Each vehicle ($i$) is described by its position $X_{i,t}$ (defined on the set of cells indexes) and velocity $V_{i,t}$ (in cells per time step). Maximal velocity $V_{max}$ is a parameter, which is assigned to the traffic stream (a set of vehicles). In order to enable appropriate modelling of signalised intersections, the saturation flow $S$ (in vehicles per hour of green time) was also taken into account as a parameter of the traffic stream. All the above quantities are expressed by triangular ordered fuzzy numbers.

### 4.1. Ordered fuzzy numbers

The concept of ordered fuzzy numbers was introduced in [30]. According to the original definition, an ordered fuzzy number ($A$) is represented by an ordered pair of continuous real functions defined on the interval [0; 1] (Fig. 4):

$$A = \langle f_A, h_A \rangle, \ f_A, h_A : [0;1] \to \mathbf{R}. \tag{5}$$

The correspondence between the ordered fuzzy numbers and the classical theory of convex fuzzy numbers was discussed in [31]. It was shown that if some specific conditions are satisfied by the pair of functions $f_A$, $h_A$ then it can be transformed into the membership function $\mu(x)$, $x \in \mathbf{R}$, which represents a convex fuzzy number in the classical sense.

The model of ordered fuzzy numbers provides a quite simple representation of non-precise information and also simple arithmetic operations. The main advantage of using the ordered fuzzy numbers is the fact that this approach eliminates several issues related to the classical fuzzy arithmetic, which is based on the so-called extension principle. In the classical approach, both the addition and subtraction operations increase fuzziness of the calculated result. In case of performing the sequences of operations repeatedly, the application of extension principle yields results with an overestimated fuzziness (imprecision) that have little potential to be useful. This fact was found to be a major obstacle impeding the construction of a fuzzy version of cellular automata model for the traffic simulation. The ordered fuzzy numbers allow the multiple operations to be performed without an excessive increase of the fuzziness.

It was assumed that only triangular fuzzy numbers will be used in the construction of a road traffic model, thus $f_A$ and $h_A$ will be affine functions. To be in agreement with the classical denotation of fuzzy sets (numbers), the independent variable of both functions will be denoted by $\mu$:

$$f_A(\mu) = a^{(1)} + \mu(a^{(2)} - a^{(1)}). \tag{6}$$
$$h_A(\mu) = a^{(3)} - \mu(a^{(3)} - a^{(2)}).$$

In the presented approach, the definition of the ordered fuzzy number was modified by introducing an interval $I_A = [a^{(0)}; a^{(4)}]$ to determine range (codomain) of the functions $f_A$ and $h_A$:

$$A = \langle f_A, h_A, I_A \rangle, \; f_A, h_A : [0;1] \to I_A. \tag{7}$$

This modification was made to enable a concise description of dependencies between fuzzy numbers representing different physical quantities, which values can vary in significantly different ranges. For further presentation of the proposed model, it will be convenient to normalise the codomain of functions $f_A$ and $h_A$ into the unit interval $[0; 1]$. Thus, the following range-normalised form of the ordered fuzzy number will be used:

$$\overline{A} = \langle \overline{f}_A, \overline{h}_A, I_A \rangle, \; \overline{f}_A, \overline{h}_A : [0;1] \to [0;1]. \tag{8}$$

where:

$$\overline{f}_A(\mu) = \frac{f_A(\mu) - a^{(0)}}{|I_A|}, \; \overline{h}_A(\mu) = \frac{h_A(\mu) - a^{(0)}}{|I_A|}. \tag{9}$$

are the range-normalised counterparts of functions $f_A$ and $h_A$ (Fig. 4).

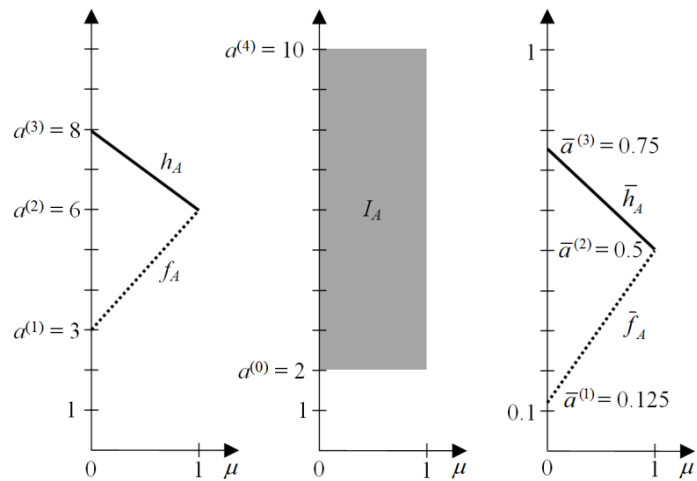

Fig. 4. Triangular ordered fuzzy number and its range-normalised counterpart

Hereinafter, all the ordered fuzzy numbers are represented by 5-tuples. The following notation is used to refer to their standard and range-normalised form respectively:

$$A = (a^{(0)}, a^{(1)}, a^{(2)}, a^{(3)}, a^{(4)}), \quad \overline{A} = (\overline{a}^{(0)}, \overline{a}^{(1)}, \overline{a}^{(2)}, \overline{a}^{(3)}, \overline{a}^{(4)}). \tag{10}$$

A definition of basic operations on triangular ordered fuzzy numbers can be now formulated using the above representation:

$$op(A, B) = C \Leftrightarrow c^{(m)} = op(a^{(m)}, b^{(m)}), \; m = 0, ..., 4, \tag{11}$$

where $A$, $B$, $C$ are ordered fuzzy numbers and $op$ stands for a particular operation: addition, subtraction, multiplication, minimum or maximum. This set of operations is sufficient for constructing a fuzzy traffic model based on the cellular automata approach.

**4.2. Traffic simulation algorithm**

In this section a traffic simulation algorithm is introduced. The most important components of this algorithm are update operations that enable the velocities and positions of vehicles to be computed at each time step of the simulation. The design of the update operations has a direct impact on the resulting saturation flow of the simulated traffic stream. The proposed algorithm allows the traffic simulation to be adjusted in order to fit a predetermined level of saturation flow, which is represented by a fuzzy number $S$.

The proposed traffic simulation algorithm (Algorithm 1) utilises a pair of deterministic rules (RL, RH) for updating the fuzzy cellular model. The rules RL and RH are used to compute positions $x_{i,t+1}^{(0)}$ and $x_{i,t+1}^{(4)}$ respectively, where $t$ denotes current time step of the simulation. For the computation of $x_{i,t+1}^{(1)}$, $x_{i,t+1}^{(2)}$ and $x_{i,t+1}^{(3)}$ one of the available rules (RL or RH) is selected at each time step. The introduced operation of rules selection allows the simulated traffic stream to reach the assumed level of saturation flow ($S$).

Algorithm 1. Traffic simulation with fuzzy cellular model

```
For t = 1 to T do
  Update traffic signals.
  For all vehicles (i = 1 to N) do
    Compute v_{i,t}^{(0)} and x_{i,t+1}^{(0)} using rule RL.
    For m = 1 to 3 do
      if  x̄_{i,t}^{(m)} ≤ α^{(m)} then compute v_{i,t}^{(m)} and x_{i,t+1}^{(m)} using rule RH,
      else compute v_{i,t}^{(m)} and x_{i,t+1}^{(m)} using rule RL.
    Compute v_{i,t}^{(4)} and x_{i,t+1}^{(4)} using rule RH.
```

As it was discussed in Section 3.1, each deterministic rule of cellular automata corresponds to a single value of the saturation flow. The value of saturation flow is determined by two parameters: maximal velocity $v_{max}$ and gap $g$ between vehicles passing through an intersection with the maximal velocity. We will denote the maximal velocity for rules RL and RH as $v_{max}^{(0)}$ and $v_{max}^{(4)}$ respectively. Similarly, the gaps for rules RL and RH will be represented by $g^{(0)}$ and $g^{(4)}$. The corresponding saturation flow values can be calculated by using Eq. (1). In the proposed algorithm the saturation flow level $s^{(0)}$, achieved for rule RL, has to be lower than the saturation flow level $s^{(4)}$ for the second rule (RH).

The pair of deterministic rules RL, RH allows us to compute an interval $I_{X_i} = [x_i^{(0)}; x_i^{(4)}]$ representing possible positions of $i$-th vehicle that correspond to the assumed interval of saturation flow values $I_S = [s^{(0)}; s^{(4)}]$. Upper chart in Fig. 5 shows positions of six vehicles, determined for the saturation flow values from interval $I_S$. Black dots represent two extreme configurations of the traffic model that are obtained by using the two different deterministic cellular automata rules. For each position $x_i^{(m)}$, such that $x_i^{(m)} \in I_{X_i}$ the following equality is satisfied:

$$\bar{x}_i^{(m)} = \bar{v}_{\max}^{(m)} = \bar{g}^{(m)}. \tag{12}$$

It should be noted here that the time indices $t$ of variables $x$ were omitted for the sake of simplicity.

On the basis of (12) we can determine the maximal velocity and gap for the vehicle position $x_i^{(m)}$:

$$v_{\max}^{(m)} = v_{\max}^{(0)} + \bar{x}_i^{(m)}(v_{\max}^{(4)} - v_{\max}^{(0)}), \tag{13}$$

$$g^{(m)} = g^{(0)} + \bar{x}_i^{(m)}(g^{(4)} - g^{(0)}). \tag{14}$$

Thus, the value of saturation flow, which corresponds to the position $x_i^{(m)}$ can be calculated as:

$$s^{(m)} = \frac{v_{\max}^{(0)} + \bar{x}_i^{(m)}(v_{\max}^{(4)} - v_{\max}^{(0)})}{g^{(0)} + \bar{x}_i^{(m)}(g^{(4)} - g^{(0)}) + 1}. \tag{15}$$

This leads to the non-linear relationships between positions $x$ of particular vehicles moving in a traffic stream and the saturation flow $s$ (Fig. 5).

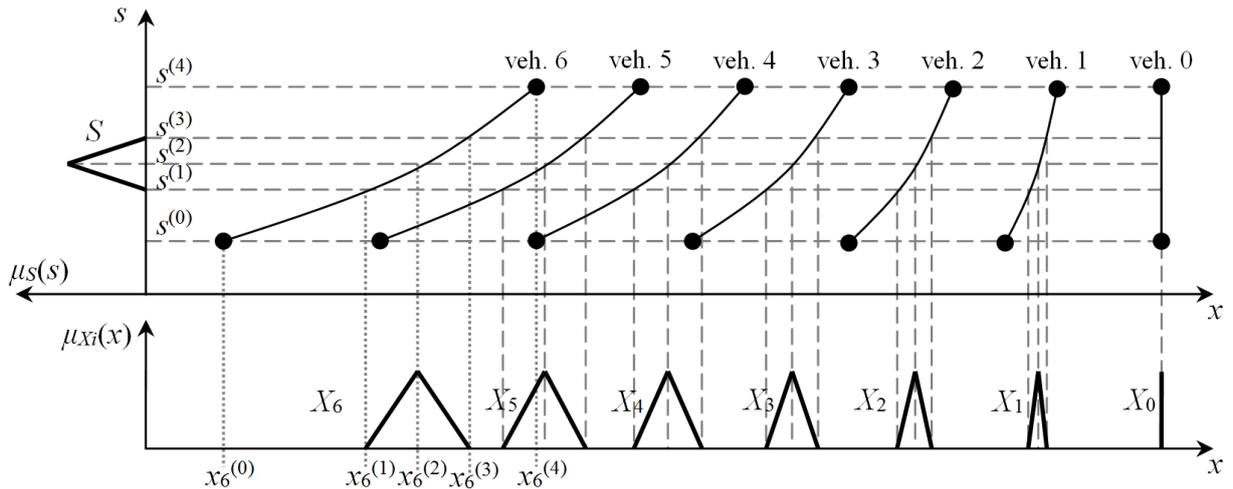

Fig. 5. Fuzzy cellular model of traffic stream: saturation flow and vehicles positions

The main aim of the introduced approach is to provide a road traffic model, which can accept an ordered fuzzy number $S$ as an input parameter specifying the level of the saturation flow. The lower chart in Fig. 5 shows vehicles positions that were obtained by taking into account the assumed fuzzy value of the saturation flow. These positions are represented by ordered fuzzy numbers $X_i$. The key insight is that by transforming the formula (15) we can determine the expected positions that correspond with the assumed level of saturation flow:

$$\bar{x}_i^{(m)} = \frac{s^{(m)}(g^{(0)} + 1) - v_{\max}^{(0)}}{(v_{\max}^{(4)} - v_{\max}^{(0)}) - s^{(m)}(g^{(4)} - g^{(0)})} = \alpha^{(m)} \quad \forall i \ \forall m. \tag{16}$$

The resulting value $\alpha^{(m)}$ is identical for all vehicles in the traffic stream and for all time steps of the simulation. It should be also noted that Eq. (16) together with the known intervals $I_{X_i}$ provides complete description of the fuzzy numbers $X_i$.

During traffic simulation the positions $x_{i,t}^{(m)}$ have to be computed as integer cell indices in accordance with the deterministic cellular automata rules. The symbol $\alpha^{(m)}$ in (16) was introduced in order to distinguish the expected positions from the current positions determined as a result of the simulation. With the above definitions, the aim of achieving the predetermined saturation flow level can be translated into the following requirement, which needs to be satisfied in the traffic simulation:

$$\left| \bar{x}_{i,t}^{(m)} - \alpha^{(m)} \right| \to \min \quad \forall i \quad \forall m \quad \forall t. \qquad (17)$$

The proposed traffic simulation algorithm copes with this minimisation problem by selecting the cellular automata rule that reduces the absolute difference defined in (17). As it was mentioned at the beginning of this section, the rule selection is executed for all vehicles at each time step of the simulation and it applies to the update of positions $x_{i,t}^{(m)}$ for $m = 1, 2, 3$. Exact definition of the rule selection operation is presented in the form of if-then statement by the pseudo-code of Algorithm 1.

Since the fuzzy cellular model is designed to be used for the simulation of signal-controlled traffic streams, it has to take into account drivers' reactions to traffic signals. The influence of traffic signalisation is simulated by introducing a set of cells in front of which the vehicles have to halt at red signals. This set will be denoted by $H$. The halt cells in $H$ correspond with the locations of stop lines. The update of a traffic signal involves the insertion of an appropriate cell index $x$ into the set $H$ when the red signal has to be activated, and removal of the index $x$ from $H$ when the red signal has to be deactivated. Thus, yellow time is considered as a part of green phase.

### 4.3. Model implementation

The following definitions provide a detailed description of the fuzzy cellular model implementation, which is based on the deterministic cellular automata rules R1 and R2 that were discussed in Section 3.1. The rules denoted by RL and RH in the simulation algorithm (Algorithm 1) are implemented here as the rules R1 and R2 respectively. The velocity of vehicle $i$ at time step $t$ is computed using the formula:

$$V_{i,t} = \min(V_{i,t-1} + 1, G_{i,t}, V_{\max}) \cdot A_{i,t}, \qquad (18)$$

where $G_{i,t}$ is the fuzzy number of free cells in front of a vehicle $i$:

$$G_{i,t} = \min({}^L G_{i,t}, {}^S G_{i,t}). \qquad (19)$$

Gap ${}^L G_{i,t}$ represents the distance between $i$-th vehicle and its lead vehicle $(i-1)$:

$$^L G_{i,t} = X_{i-1,t} - X_{i,t} - 1. \qquad (20)$$

If there is no lead vehicle in front of the vehicle $i$ then $G_{i,t}$ is assumed to be equal to $V_{\max}$.

The variable ${}^S G_{i,t}$ describes distance of the $i$-th vehicle to the nearest red signal (i.e. the halt cell $x \in H$):

$$^S g_{i,t}^{(m)} = \min\{x - x_{i,t}^{(m)} - 1 : x > x_{i,t}^{(m)} \wedge x \in H\}. \qquad (21)$$

If there is no halt cell $x$ satisfying the condition in (21) then ${}^S g_{i,t}^{(m)}$ is assumed to be equal to ${}^L g_{i,t}^{(m)}$.

The variables $V_{i,t}$, $G_{i,t}$, ${}^L G_{i,t}$, ${}^S G_{i,t}$, $X_{i,t}$, and $V_{\max}$ are triangular ordered fuzzy numbers. Thus, the calculations are performed according to the definition (11). The subtraction of 1 in (20) is handled by interpreting the scalar value as an ordered fuzzy number, which can be represented by 5-tuple having all the components equal to one. $A_{i,t}$ is a 5-tuple of binary values:

$$a_{i,t}^{(m)} = \begin{cases} 0, & r_{i,t}^{(m)} = \mathrm{RL} \wedge v_{i,t-1}^{(m)} = 0 \wedge g_{i,t}^{(m)} = 1, \\ 1, & \text{else,} \end{cases} \qquad (22)$$

where $m = 0, ..., 5$. Note that if $a_{i,t}^{(m)} = 0$ then $v_{i,t}^{(m)}$ is computed by using the rule RL (i.e. R1). The multiplication by the binary tuple $A_{i,t}$ in (18) is performed similarly to the operations on ordered fuzzy numbers, according to the definition (11).

After the determination of velocities for all vehicles, their positions are updated as follows:

$$X_{i,t+1} = X_{i,t} + V_{i,t} \cdot B_{i,t},  \qquad (23)$$

where $B_{i,t}$ is a 5-tuple of binary values:

$$b_{i,t}^{(m)} = \begin{cases} 0, & r_{i,t}^{(m)} = \text{RH} \land v_{i,t-1}^{(m)} = 0 \land g_{i,t}^{(m)} = 1, \\ 1, & \text{else.} \end{cases} \qquad (24)$$

In the above formula, $b_{i,t}^{(m)}$ takes value 0 when the rule RH (i.e. R2) is used for the computation of $x_{i,t+1}^{(m)}$.

The variables $r_{i,t}^{(m)}$ in equations (22) and (24) are components of another 5-tuple ($R_{i,t}$), which is introduced to describe the current selection of rules:

$$r_{i,t}^{(m)} = \begin{cases} \text{RH}, & \bar{x}_{i,t}^{(m)} \leq \alpha^{(m)} \\ \text{RL}, & \text{else} \end{cases},\ m = 1,...,3,\ r_{i,t}^{(0)} = \text{RL},\ r_{i,t}^{(4)} = \text{RH}. \qquad (25)$$

It should be noted here that the above definition of $R_{i,t}$ is consistent with the rule selection operation that was introduced for the fuzzy cellular model in Section 4.2.

## 5. Comparison with Nagel-Schreckenberg cellular automata model

### 5.1. Simulation results

This section presents some results of traffic simulations in a signalised arterial. The results of a simulation performed with the fuzzy cellular model are compared against those obtained by using the Nagel-Schreckenberg (NaSch) stochastic cellular automata model. For the purpose of this experiment the fuzzy cellular model was implemented according to the definitions given in Section 4.

The parameters of the fuzzy cellular model were adjusted to match the following settings of the stochastic NaSch model: time step 1 s, cell length 7.5 m deceleration probability $p = 0.2$ and maximal velocity $v_{\max} = 2$ cells per time step. Similarly, for the fuzzy cellular model the maximal velocity of vehicles was assumed as two cells per time step: $v_{\max}^{(m)} = 2\ \forall m$. Thus, due to the application of deterministic rules R1 and R2, the resulting interval of saturation flow values for the fuzzy cellular model is $I_S = [1440, 1800]$ (in vehicles per hour of green time).

The saturation flow volume in the fuzzy cellular model is defined by the ordered fuzzy number $S$. This parameter was estimated in order to reproduce the distribution of saturation flow rates observed in the NaSch model. Fig. 6 shows a histogram of the saturation flow values that were obtained for the NaSch model during 500 runs of a traffic simulation at signalised intersection. The simulation period was 3600 seconds for each run. The experimental data presented in Fig. 6 were further used for the determination of the parameter $S$. The values of $s^{(1)}$, $s^{(2)}$ and $s^{(3)}$ were set respectively as the 5-th percentile, median and 95-th percentile of the saturation flow rates distribution. As a result the following fuzzy number was obtained, which describes the saturation flow in vehicles per hour of green time: $S = (1440, 1503, 1575, 1638, 1800)$. The parameters $\alpha^{(1)} = 0.21$, $\alpha^{(2)} = 0.43$, $\alpha^{(3)} = 0.60$ of the fuzzy cellular model were calculated for the above level of saturation flow according to Eq. (16).

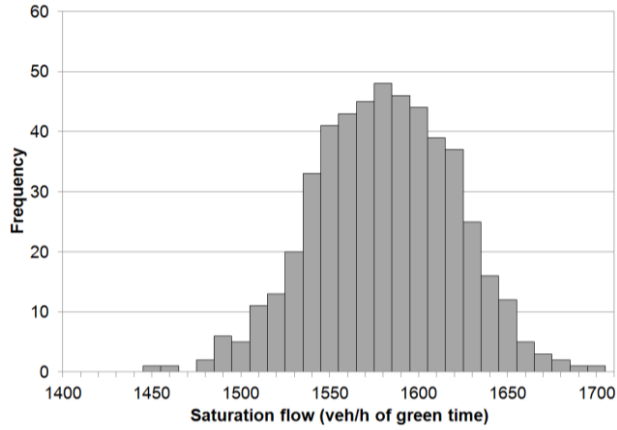

Fig. 6. Histogram of saturation flow rates for NaSch model

The cell length for the fuzzy cellular model was determined on the basis of free-flow velocity comparison. For low traffic densities in NaSch model, vehicles move with a free-flow velocity, which can be calculated in cells per time step using the formula: $v_f = v_{max} - p$. In case of the fuzzy cellular model the free-flow velocity is equal to $V_{max}$ (note that $V_{max}$ is an ordered fuzzy number). The time step for both models corresponds to one second. Taking into account the parameters that were determined above, the cell length for fuzzy cellular model can be calculated as $7.5 \cdot (v_{max} - p) / v_{max}^{(m)} = 6.75$ m. Thus, in both models the free-flow velocity is equal to 13.5 m/s (48.6 km/h).

Simulations of a traffic in signalised one-way arterial road were performed in order to compare the introduced fuzzy cellular model and the NaSch cellular automata. Parameters of both models were set as discussed above. One-lane road was considered in this study because the analysis is focused on the detailed properties of the models that are related to single lane traffic streams. It was assumed that three signalised intersections are located at the modelled road. Total length of the road is 3 km and the distances between intersections are equal to 750 m.

The initial conditions of the traffic simulations are determined by a single queue length parameter (i.e. number of vehicles waiting in a queue at an intersection). At the beginning of each simulation queues of equal length are formed for all three intersections. Additionally, the last vehicle is always inserted into the first cell of the modelled road.

Fig. 7 presents trajectories of the last vehicle in time-space diagrams. Black dashed and doted lines show trajectories that were determined by using the fuzzy cellular model (FCM). The grey colour indicates trajectories obtained for the NaSch model during 500 runs of the traffic simulation. Black horizontal bars correspond to red time intervals at the intersections. Traffic signal timings are similar for all simulated intersections. The results presented in Fig. 7 a) were obtained for signal cycle time of 60 s and green phase of 30 s. For the second case (Fig. 7 b) the cycle time was 90 s and green phase was 45 s.

Dependency between the queue length parameter and the travel time of the last vehicle is illustrated in Fig. 8. The travel time is defined as time required for the last vehicle to pass a stop line at the third intersection. The results of travel time estimation are compared for the two analysed models. In case of the NaSch model application, the percentiles of travel time distribution are determined on the basis of 500 simulation runs for each queue length. Using fuzzy cellular model, the travel time is determined in single simulation run as an ordered fuzzy number $\Theta$ according to the following equation:

$$\theta^{(m)} = \min\{t : x_{l,t}^{(m)} > 333\}, \qquad (26)$$

where *l* is an index of the last vehicle. The cell number 333 corresponds to location of the stop line at third intersection.

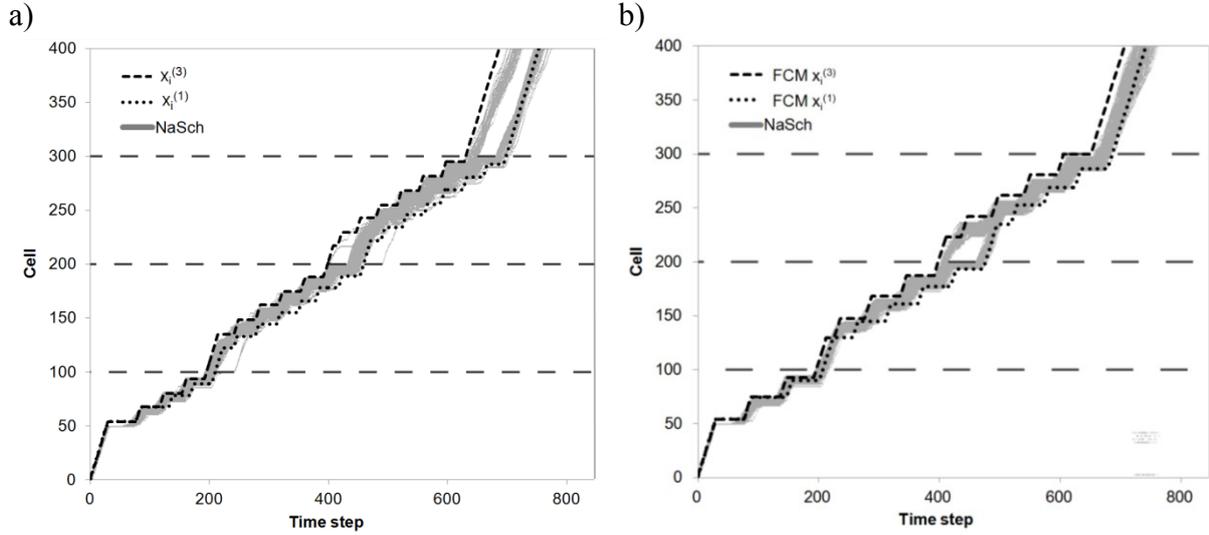

Fig. 7. Trajectory of the last vehicle for two signal cycle times: a) 60 s b) 90 s

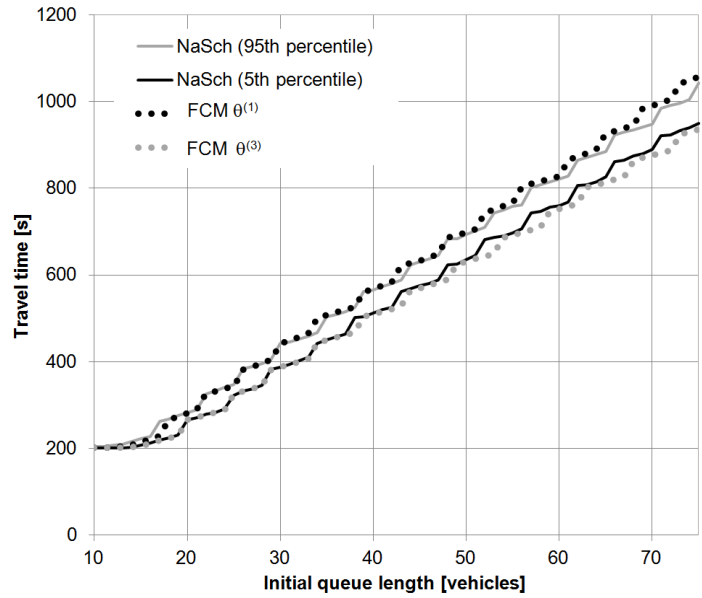

Fig. 8. Travel time of the last vehicle vs. initial queue length

Plots in Fig. 9 show number of vehicles on the modelled road (upstream of the third intersection) for successive time steps of the traffic simulation. Note that the vehicles are inserted into the modelled road only at the beginning of simulation. The number of vehicles decreases during simulation as subsequent vehicles pass over the stop line of the third intersection. The initial number of vehicles is determined by the queue length parameter. The results presented in Fig. 9 a) and b) were obtained for the queue length of 30 and 70 vehicles respectively. The results for NaSch model were estimated after 500 runs of the traffic simulation. In the fuzzy cellular model, the number of vehicles at time step *t* is directly calculated for a single simulation run as an ordered fuzzy number $N_t$:

$$n_t^{(m)} = \left| \{i : x_{i,t}^{(m)} \leq 333\} \right|, \quad (27)$$

where |.| denotes the cardinality of a set.

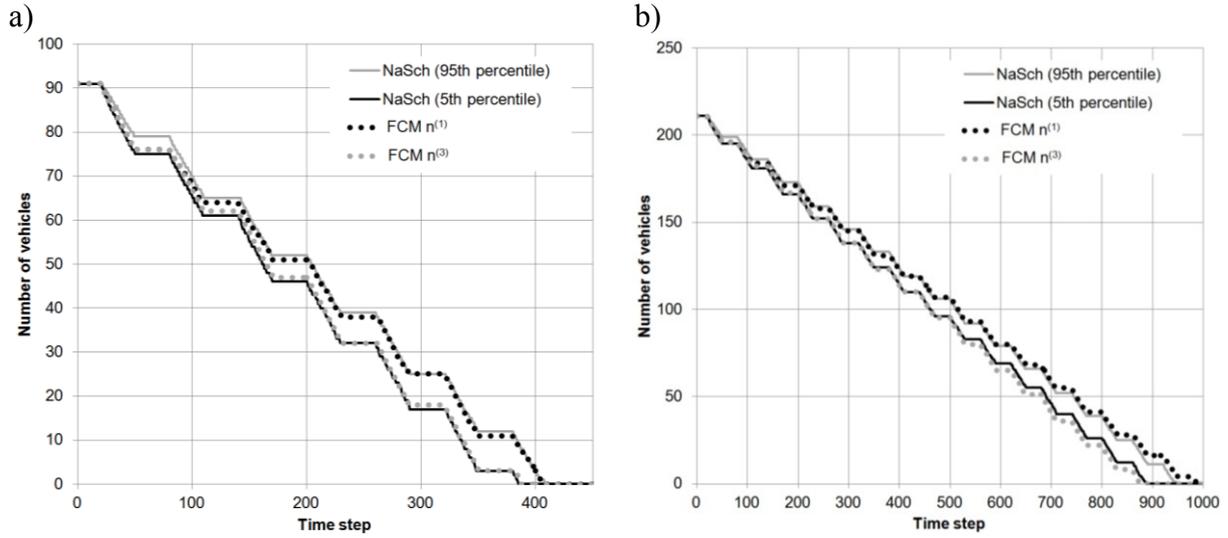

Fig. 9. Number of vehicles on the modelled road for two initial queue lengths: a) 30 b) 70

The results shown in Fig. 8 and Fig. 9 were obtained for the traffic signal cycle of 60 s with green phase of 30 s. The above signal timing parameters were used for all three intersections.

As it can be observed in the above results, the evolution of the simulated traffic stream is very similar for both models. This fact proves that the proposed fuzzy cellular model can be appropriately calibrated to reproduce the traffic stream behaviour for a given distribution of saturation flows rates. The experiments have shown that the accuracy of traffic simulation is similar for both considered models. However, the fuzzy cellular model avoids the aforementioned disadvantages of the cellular automata (Section 3). Firstly, the proposed model can be precisely calibrated by adjusting its parameters. Moreover, the uncertainty of model parameters can be taken into account as the parameters are represented by fuzzy numbers. Secondly, the fuzzy cellular model does not need multiple simulations because it uses the fuzzy numbers to estimate the distributions of traffic performance measures (travel time, the number of vehicles in a given region, delays, queue lengths, etc.) during a single run of the traffic simulation.

## 5.2. Computational cost

The implementation of the NaSch model requires multiple traffic simulation runs (see Algorithm 2). At each run, the simulation results have to be stored. After $K$ runs, the stored results are used to calculate distributions of the traffic performance measures. The number of simulation runs $K$ has to be appropriately high in order to obtain meaningful estimates. For the experiments presented in this section the number of runs $K$ was 500.

The stochastic rule of NaSch cellular automata was decomposed into two deterministic rules, denoted by NSH and NSL. The NSH rule is consistent with the R2 rule, which was defined in Section 3.1. It corresponds to the NaSch rule with parameter $p = 0$. The NSL rule reflects the operation of the NaSch rule for $p = 1$. Thus, the velocity in the NSL rule is calculated according to the following formula:

$$v_{i,t} = \max(0, \min(v_{i,t-1}+1, g_{i,t}, v_{max})-1). \tag{28}$$

The randomisation step of the NaSch model was implemented in the simulation algorithm by introducing a selection of the deterministic rule (NSL or NSH). The selection is based on a random number $\xi \in [0;1)$, which is drawn from a uniform distribution. This description of the traffic simulation algorithm enables its comparison with the algorithm proposed for the fuzzy cellular model (Algorithm 1).

Algorithm 2. Traffic simulation with the NaSch model

```
For simulation run 1 to K do
  For t = 1 to T do
    Update traffic signals.
    For all vehicles (i = 1 to N) do
      Generate random number ξ.
      If ξ < p then compute v_{i,t} and x_{i,t+1} using rule NSL,
      else compute v_{i,t} and x_{i,t+1} using rule NSH.
    Store simulation results.
```

Let us assume that the basic operation in the traffic simulation algorithm is the execution of deterministic cellular automata rule i.e. the computation of the position and velocity for a single vehicle. The number of basic operations performed during the traffic simulation can be determined for both compared models by analysing the pseudo-code of Algorithm 1 and Algorithm 2. The traffic simulation with the NaSch model requires $K \cdot T \cdot N$ basic operations whereas during the simulation with the fuzzy cellular model the basic operation is executed $5 \cdot T \cdot N$ times. It was assumed that the number of vehicles $N$ is constant in the analysed simulation period. The computational cost of traffic simulation is considerably reduced for the fuzzy cellular model because the number of simulation runs $K$ is always much greater than 5 (usually amounts to several hundred runs). Moreover, the traffic simulation with the fuzzy cellular model does not need to store partial results, thus it requires less memory space than the simulation with the NaSch cellular automata.

## 6. Conclusions

The fuzzy cellular model of signal controlled traffic stream was proposed by combining cellular automata with fuzzy calculus. The presented approach benefits from advantages of the cellular automata models and eliminates the main drawbacks that have impeded their applications in traffic control systems. Parameters of the fuzzy cellular model enable a simple calibration and allow the traffic simulation to reflect predetermined saturation flow rates. The fuzzy numbers are used in order to describe the uncertainty and precision of the simulation inputs and outputs. Thus, the imprecise traffic data can be utilised in the proposed modelling approach for the estimation of traffic performance [32]. The experiments reported in this paper show that the traffic simulations with the fuzzy cellular model are consistent with those performed by stochastic cellular automata. It was also demonstrated that the application of the introduced model considerably reduces the computational cost of traffic simulation. These findings are of vital importance for real-time applications of microscopic models in the road traffic control.